\documentclass[aps,prl,reprint,superscriptaddress,twocolumn,showpacs,superscriptaddress]{revtex4-1}
\usepackage[latin1]{inputenc}
\usepackage{amsmath}
\usepackage{amsfonts}
\usepackage{amssymb}
\usepackage{lmodern,dsfont}
\usepackage{graphicx}
\usepackage[usenames,dvipsnames]{xcolor}

\newcommand{\kb}{k_\text{B}}

\begin{document}

\title{Underdamped stochastic heat engine at maximum efficiency}

\author{Andreas Dechant}
\affiliation{Department of Physics, Friedrich-Alexander-Universit\"at Erlangen-N\"urnberg, 91058 Erlangen, Germany}
\author{Nikolai Kiesel}
\affiliation{Vienna Center for Quantum Science and Technology (VCQ), Faculty of Physics, University of Vienna, A-1090 Vienna, Austria}
\author{Eric Lutz}
\affiliation{Department of Physics, Friedrich-Alexander-Universit\"at Erlangen-N\"urnberg, 91058 Erlangen, Germany}

\begin{abstract}
We investigate the performance of an underdamped  stochastic heat engine for a time-dependent harmonic oscillator. 
We analytically determine the optimal  protocol that maximizes the efficiency at fixed power. The maximum efficiency reduces to the Curzon-Ahlborn formula at maximum power and the Carnot formula at zero power. We further establish that the efficiency at maximum power is universally given by the Curzon-Ahlborn efficiency in the weakly damped regime. Finally, we show that even small deviations from operation at maximum power may result in a significantly increased efficiency.
\end{abstract}


\maketitle

The universal upper bound on the efficiency of any heat engine operating between two equilibrium   baths at temperatures $T_c$ and $T_h$ ($T_c<T_h$) is given by the Carnot formula, $\eta_\text{C} = 1 - T_c/T_h$ \cite{cen01}. The Carnot efficiency is however only reachable in the reversible limit, for ideal engines that run infinitely slowly and thus have zero power. It is hence customary to consider the efficiency at maximum power $\eta^*$, popularized by Curzon and Ahlborn \cite{cur75}, which yields a better estimate for the performance of real engines. For low-dissipation machines, generic bounds on $\eta^*$ have been obtained in Refs.~\cite{esp09,esp10}. They reduce to the Curzon-Ahlborn efficiency, $\eta_\text{CA}=1-\sqrt{T_c/T_h}$,  for symmetric coupling to  the cold and hot baths. The existence of these general bounds is of fundamental importance. Yet, they do not address the crucial question of how to actually reach them, as they do not provide information about optimal engine cycles. It has  recently been shown that seeking the efficiency at maximum power may not always be the best strategy \cite{whi14,whi15}. To achieve engines with high efficiency and high power one may rather want to maximize the efficiency at a given power. A motor working at maximum efficiency may, for instance, attain higher efficiency and power than motors working at their maximum power. Devices that supply a fixed power, with the highest possible efficiency, are moreover often required in practical engineering applications. The efficiency at given power output, as opposed to  maximum power output, additionally  offers   a direct method to compare the performance of different engines.

In this paper, we compute the maximum efficiency at fixed power $\tilde \eta$ for an underdamped harmonic stochastic heat engine and  determine the corresponding optimal driving protocols. Such a system may be  viewed as a minimal model for a microscopic piston engine: 
Increasing and decreasing the strength of the harmonic potential corresponds to the compression and expansion phase of the cycle, whereas heating and cooling is accomplished by changing the temperature of the surrounding medium. In the overdamped limit that describes strong particle-medium coupling and fast thermalization, the optimization problem has been solved by Schmiedl and Seifert \cite{sch07}: they have calculated the optimal protocols that lead to maximal power output and derived the corresponding efficiency at maximum power. Interestingly, they found that $\eta^*$ may surpass the Curzon-Ahlborn efficiency $\eta_{CA}$. Experimental realizations of microscopic Brownian engines   in  colloidal systems have been reported in Refs.~\cite{bli12,mar15}. A proposal to implement a stochastic heat engine using  an optically  levitated nanoparticle in the opposite limit of weak damping and slow thermalization has  recently been put forward in Ref.~\cite{dec15}. Optimization in this underdamped regime is significantly more difficult, owing to the coupling between position and velocity dynamics  \cite{gom08,mur14a,mur14}. It has for this reason been little explored so far. In the following, we optimize the heat engine for maximum efficiency at constant power in the underdamped limit.  We  compute  the optimal frequency driving protocols and show that they exhibit discontinuities at the change from the hot to the cold bath and vice versa. We  obtain an explicit expression for the maximum efficiency at given power that contains both the Curzon-Ahlborn efficiency at maximum power and the Carnot efficiency at zero power as limiting cases. We  further demonstrate that the Curzon-Ahlborn formula provides a universal upper bound to the efficiency at maximum power in the underdamped regime, in contrast to the overdamped limit. We finally  show that it is possible to considerably enhance the efficiency of the  engine by sacrificing relatively little power.

\textit{Brownian heat engine.} 
We consider  a stochastic heat engine consisting of a harmonically bound  particle coupled to a heat bath.
We assume that its dynamics is described by the Langevin equations,
\begin{subequations}
 \label{langevin}
\begin{align}
m \dot{v} &= - m \gamma v - m \omega^2 x + \sqrt{2 m \gamma \kb T}\, \xi, \\
\dot{x} &= v.
\end{align} 
\end{subequations}
Here $m$, $v$ and $x$ are the respective mass, velocity and position of the Brownian particle.
The heat bath is characterized by the damping coefficient $\gamma$ and the temperature $T$.
In the following, we set $\kb = 1$ for simplicity.
The quantity $\xi(t)$ is a centered Gaussian white noise force with $\langle \xi(t) \xi(t') \rangle = \delta(t-t')$.
In order to operate this system as a heat engine, we cyclically vary the temperature $T$ and the frequency $\omega$ of the harmonic potential  as a function of time.
The harmonic confinement may  be realized, for example, via optical tweezers \cite{bli12,mar15} or via a standing light wave in a cavity \cite{dec15}.
Since Eq.~\eqref{langevin} is linear in $x$ and $v$, the phase-space distribution $P(x,v,t)$ will always be Gaussian, provided that the initial state $P(x,v,0)$ is Gaussian.
We may accordingly  describe the system with an equivalent set of deterministic equations of motion for the variances, $\sigma_x = \langle x^2 \rangle$ and $\sigma_v = \langle v^2 \rangle$,
\begin{subequations} \label{variances}
\begin{align}
\dot{\sigma}_v + 2 \gamma \sigma_v + \omega^2 \dot{\sigma}_x  &=  \frac{2 \gamma T}{m} \label{variances-a},\\
\ddot{\sigma}_x - 2 \sigma_v + 2 \omega^2 \sigma_x + \gamma \dot{\sigma}_x &= 0 . \label{variances-b}
\end{align} 
\end{subequations}
We will  discuss two  different regimes for the dynamics of the particle \cite{ris89}.
In the overdamped limit, $\gamma \gg \omega$, the velocity distribution relaxes to its thermal equilibrium form on time scales that are short compared to the slow spatial motion in the potential;
it  may thus be assumed to be always thermal, $P_v(v) = Z^{-1} \exp(-m v^2/(2 T))$.
The velocity variable may then be eliminated  by setting $\sigma_v = T/m$ in Eq.~\eqref{variances-b}, leading to  a single equation for $\sigma_x$:
\begin{align}
\dot{\sigma}_x + \frac{2 \omega^2}{\gamma} \sigma_x = \frac{2 T}{m \gamma} \label{variance_x}.
\end{align}
We have here assumed that the second derivative of $\sigma_x$ is small in the long-time limit.
In the opposite, underdamped limit, $\omega \gg \gamma$, the particle completes many oscillations in the potential before its energy is significantly changed through the interaction with the bath.
By virtue of the virial theorem, we may hence take kinetic and potential energy to be equal, $m \sigma_v/2 = m \omega^2 \sigma_x/2$.
Replacing $\sigma_x$ by $\sigma_v/\omega^2$ in Eq.~\eqref{variances-a} and keeping in mind that $\omega$ may be time-dependent, we find to leading order in $\gamma/\omega$,
\begin{align}
\dot{\sigma}_v + \Big(\gamma - \frac{\dot{\omega}}{\omega} \Big)\sigma_v = \frac{\gamma T}{m}  \label{variance_v}.
\end{align}
Whereas the system adjusts instantly to the bath in the overdamped limit, it adjusts immediately to changes of the external potential  in the underdamped limit.

During a cycle of the heat engine, the particle is coupled to the hot bath at temperature $T_h$ for a time $\tau_h$, and then to the cold bath at temperature $T_c < T_h$ for a time $\tau_c$.
We  assume the switch between the baths to be quasi-instantaneous, as in the experiment  \cite{bli12,mar15} .
We further allow for different bath couplings $\gamma_h$ and $\gamma_c$.
During this procedure,  the frequency $\omega$ is varied with time to compress or expand the system.
After time $\tau_h + \tau_c$, the particle is again coupled to the hot bath and  the cycle starts anew.
In the underdamped regime, the heat exchanged with any of the two  baths $(i=h,c)$ is  given by \cite{sek10,sei12},
\begin{align}
 Q_i = \gamma_i \Big[ \tau_i T_i - m \int_{0}^{\tau_i} \text{d}t \ \sigma_v(t) \Big], \label{heat-underdamped}
\end{align}
where $\sigma_v(t)$ is the solution of Eq.~\eqref{variance_v} with the appropriate boundary conditions.
Here we define  heat such that it is positive if the particle absorbs energy from the bath.
The energy change, $\Delta U =  Q +  W$, vanishes over one cycle and  the extracted work is equal to the total  heat, $- W = Q_h + Q_c$.
As a result,  power output $P$ and efficiency $\eta$ of the Brownian heat engine may be written as,
\begin{equation}
\label{6}
P = \frac{ Q_h +  Q_c}{\tau_h + \tau_c},\qquad \eta = \frac{Q_h +  Q_c}{Q_h}. 
\end{equation}

\textit{Optimal protocols.} 
We now fix the power at a given value, $P = P_0$,  and seek  the optimal protocol $\omega(t)$ that maximizes the efficiency $\eta$.
According to Eq.~\eqref{6}, we need to simultaneously minimize $-Q_c$, the heat dissipated to the cold bath, and maximize $Q_h$, the heat uptake from the hot bath.
In a first step, we fix the coupling times $\tau_h$ and $\tau_c$.
In the underdamped regime, we have to maximize Eq.~\eqref{heat-underdamped} with the constraint \eqref{variance_v} imposed on $\sigma_v$.
The effective Lagrangian for this optimization problem is,
\begin{align}
\!\mathcal{L}(\sigma_v, \dot{\sigma}_v, \omega, \dot{\omega}, \alpha)\! =\! \sigma_v + \alpha \Big[\dot{\sigma}_v + \Big(\gamma - \frac{\dot{\omega}}{\omega} \Big)\sigma_v - \frac{\gamma T}{m}\Big].
\end{align}
The corresponding Euler-Lagrange equations read,
\begin{subequations} \label{EL-under}
\begin{align}
1 + \alpha \Big(\gamma - \frac{\dot{\omega}}{\omega} \Big) - \dot{\alpha} &= 0 \label{EL-under1} \\
\frac{1}{\omega} \frac{\text{d}}{\text{d}t} (\alpha \sigma_v) &= 0 \label{EL-under2}\\
\dot{\sigma}_v + \Big(\gamma - \frac{\dot{\omega}}{\omega} \Big)\sigma_v - \frac{\gamma T}{m} &= 0. \label{EL-under3}
\end{align} 
\end{subequations}
We provide a detailed solution of the above equations in the Supplemental Material. In particular, we establish  that a nontrivial solution of the optimization problem  requires discontinuities in both $\omega(t)$ and $\sigma_v(t)$ at the transitions between cold and hot baths.
We note that in the underdamped limit, a discontinuous variance $\sigma_v$ is permitted, since we assume that the system instantly adjusts to the potential.
From the condition $\sigma_{+} \omega_{-} = \sigma_{-} \omega_{+}$ at $t= \tau_h$ and a similar one at $t = \tau_h + \tau_c$, we find,
\begin{subequations} \label{underdamped-solution}
\begin{align}
&\omega(t) = \omega_c \left\lbrace \begin{array}{ll}
\frac{\sigma_h}{\sigma_c} e^{\gamma_h\big(1-\frac{T_h}{m \sigma_h}\big) (t-\tau_h)} &\text{for} \; 0 < t < \tau_h, \\[2ex]
 e^{\gamma_c\big(1-\frac{T_c}{m \sigma_c}\big) (t-\tau_h)} & \text{for} \; 0 < t - \tau_h < \tau_c . \label{protocol-under-2}
\end{array} \right. \\
&\gamma_h \tau_h \Big(1 - \frac{T_h}{m \sigma_h} \Big) + \gamma_c \tau_c \Big(1 - \frac{T_c}{m \sigma_c} \Big) = 0 . \label{sigma-condition}
\end{align} 
\end{subequations}
The above equations completely specify the optimal frequency protocol $\omega(t)$ that leads to extremal values of $Q_h$ and $Q_c$ for fixed $\tau_c$ and $\tau_h$ and arbitrary $\omega_c$. They constitute the first main result of this paper.

\begin{figure}[t]
\includegraphics[width=0.48\textwidth]{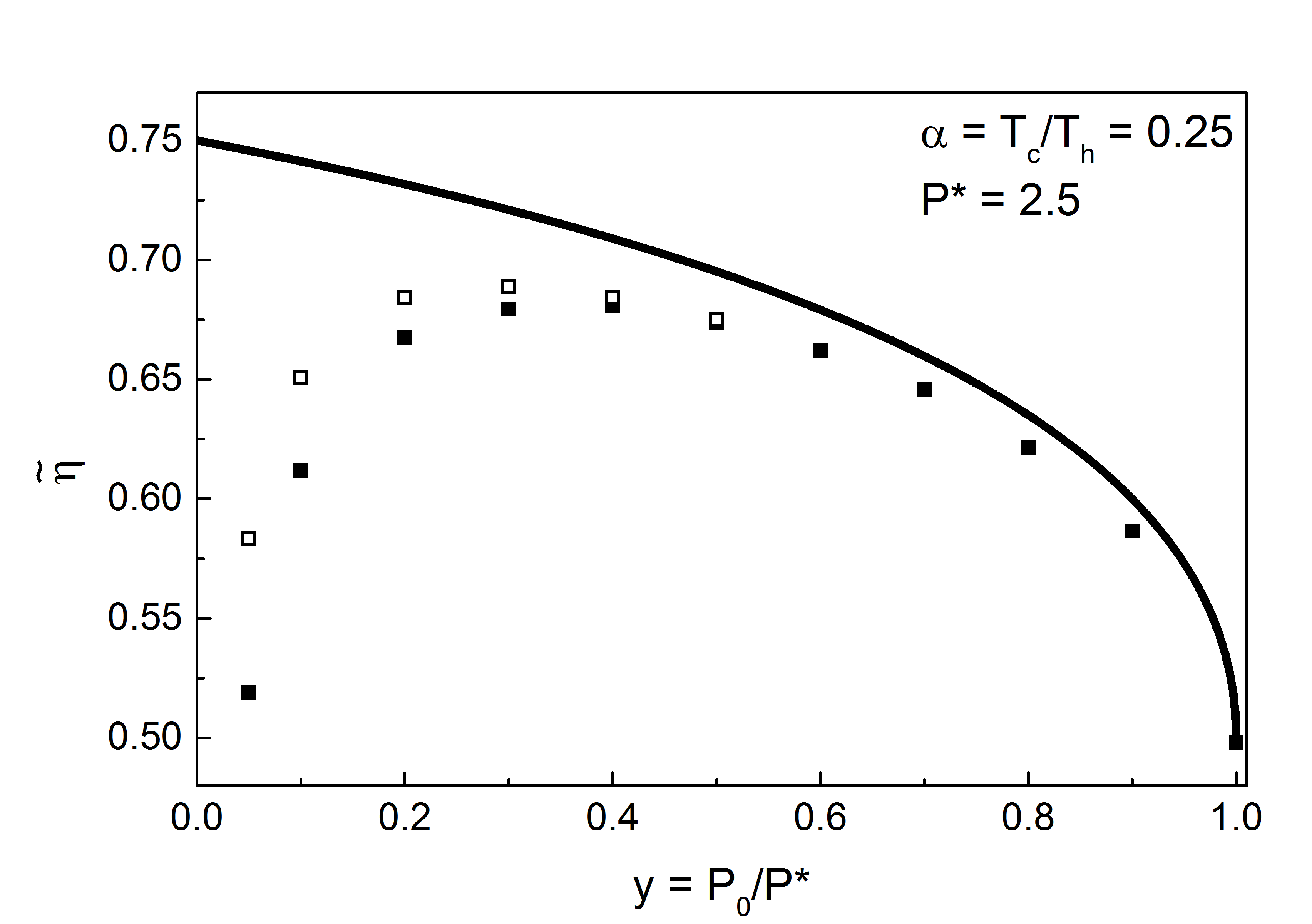}
\caption{Maximum efficiency $\tilde{\eta}$ at constant power $P_0 = y P^{*}$ as a function of $y$ for a temperature ratio of $\alpha=T_c/T_h = 0.25$. The line is the analytic underdamped result \eqref{max-efficiency-under}, the dots are obtained by numerically solving the exact dynamics \eqref{variances} for $\omega/\gamma = 1000$ (full symbols) and $\omega/\gamma = 2000$ (empty symbols). At maximum power, $P_0 = P^*$, the efficiency reduces to the Curzon-Ahlborn formula $\eta_{CA}=1-\sqrt{T_c/T_h}$. Even slightly reducing the power yields a significantly enhanced efficiency. \label{fig:maxeff}}
\end{figure}

\textit{Maximum efficiency at fixed power.} 
In order to determine the maximum efficiency of the stochastic   engine, $\eta = 1 + [\gamma_c \tau_c (T_c - m \sigma_c)]/[\gamma_h \tau_h (T_h - m \sigma_h)]$, we eliminate $\sigma_h$ and $\sigma_c$ using Eqs.~\eqref{6}, with $P = P_0$, and Eq.~\eqref{sigma-condition}.
Maximizing the efficiency $\eta$ with respect to $\tau_h$, $\partial_{\tau_h} \eta = 0$, we obtain a relation between the two coupling times,
\begin{align}
\frac{\tau_h}{\tau_c} &= \sqrt{\frac{\gamma_c}{\gamma_h}} . \label{coupling-ratio}
\end{align}
Using Eq.~\eqref{coupling-ratio}, we arrive after some algebra at an explicit expression for the maximum efficiency $\tilde \eta$,
\begin{align}
\tilde{\eta} &= \Bigg[2 y (1 - \sqrt{\alpha})^2 \Bigg] \Bigg/ \Bigg[1-\alpha+ y (1 - \sqrt{\alpha})^2 \nonumber \\
& \qquad - \sqrt{(1-y)\Big(y(1-\sqrt{\alpha})^4 + (1-\alpha)^2 \Big)} \Bigg], \label{max-efficiency-under}
\end{align}
where we have defined $\alpha = T_c/T_h$ and $y = P_0/P^{*}$ .
The maximum power $P^{*}$ is given by,
\begin{align}
P^{*} &= \frac{\gamma_h \gamma_c}{(\sqrt{\gamma_h}+\sqrt{\gamma_c})^2} \Big[ T_h + T_c - 2 \sqrt{T_h T_c} \Big]. \label{max-power-under}
\end{align}
Equation \eqref{max-efficiency-under} is our second main result. It expresses the maximum efficiency $\tilde \eta$ as a function of the output power $P_0$ (relative to the maximal value $P^*$)  and the temperature ratio $\alpha$.
For vanishing  power, $y = 0$, $\tilde \eta$ reduces to the Carnot efficiency, $\eta_\text{C} = 1 - T_c/T_h$.
On the other hand, for maximal  power, $y = 1$, we obtain the efficiency, $
\eta^{*} = 1 - \sqrt{{T_c}/{T_h}}$. 
This is precisely the Curzon-Ahlborn efficiency $\eta_{CA}$ found for an endoreversible heat engine with linear heat conduction \cite{cur75}.
The underdamped Brownian heat engine is an ideal realization of this type of engine:
Due to the small dissipation, the engine is endoreversible, and the expansion for small $\gamma$ ensures linear coupling to the heat bath. We emphasize that this result is independent of the details of $\gamma$ in the small $\gamma$ limit \cite{com}. The Curzon-Ahlborn formula therefore provides a generic  upper bound to the efficiency at maximum power in the underdamped regime.
For intermediate power, $0 < P_0 < P^{*}$, Eq.~\eqref{max-efficiency-under} interpolates between the Carnot and Curzon-Ahlborn efficiencies, see Fig.~\ref{fig:maxeff}.
Close to maximum power, we may expand around $y=1$ to find $\tilde{\eta} \simeq \eta^*(1 + (T_c/T_h)^{1/4} \sqrt{1-y})$.
The efficiency is thus nonanalytic with a diverging slope at this point.
The physical consequence of this behavior is that even slightly reducing the power below the maximal value will yield disproportionately large gains in efficiency.
This effect is most pronounced for small temperature differences, $T_c/T_h \approx 1$.
This situation, where the overall efficiency is small, is often encountered in practice, since large temperature gradients are hard to maintain in microscopic systems.
In this regime, it is thus possible to significantly increase the efficiency of the heat engine by sacrificing a relatively small amount of power.

\begin{figure}[t]
\includegraphics[width=0.48\textwidth]{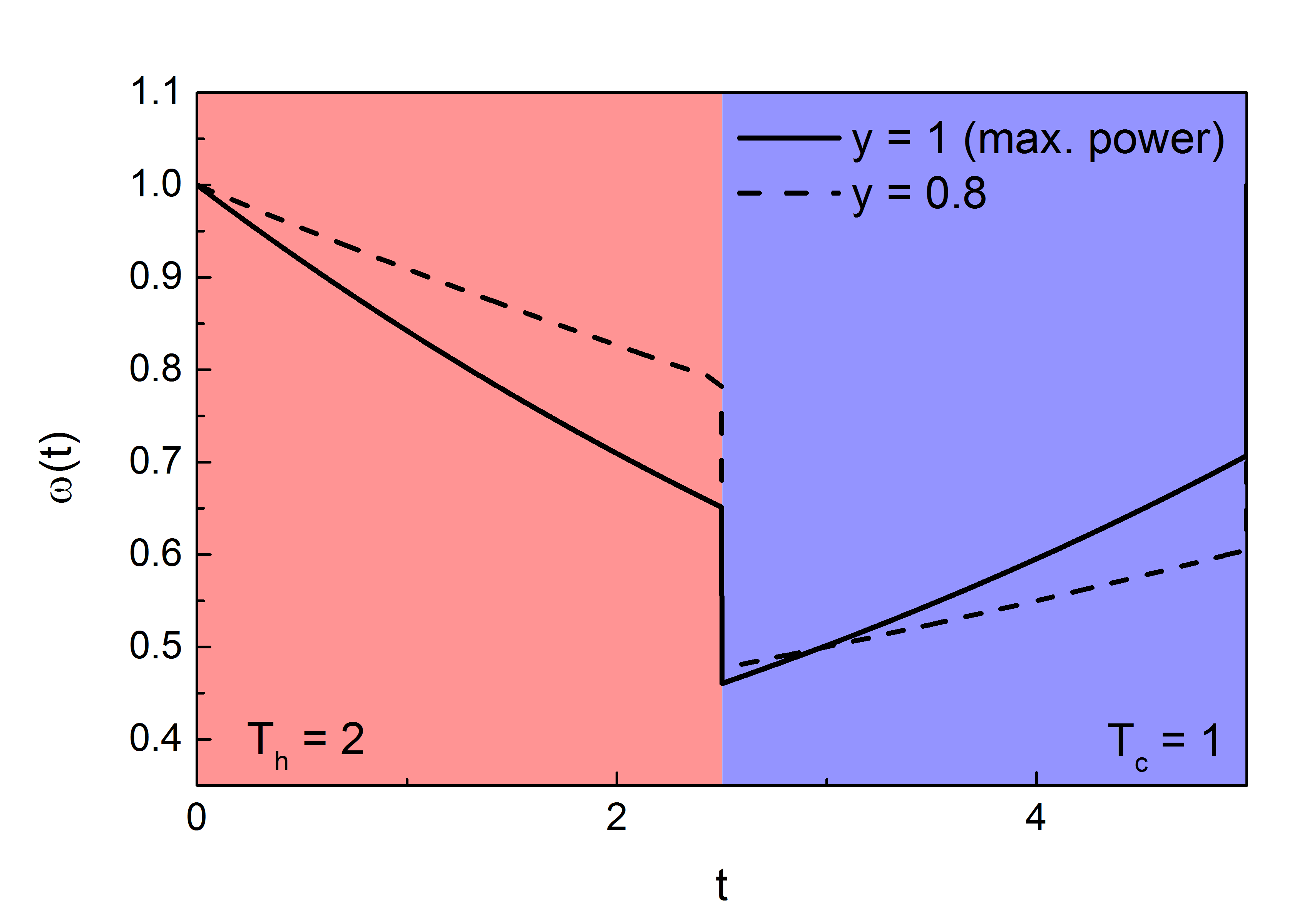}
\caption{(color online) Optimal frequency protocols for an underdamped stochastic heat engine. The parameters are $T_h/T_c = 2$ and symmetric coupling $\gamma_h = \gamma_c$. The red (blue) shaded background shows the coupling to the hot (cold) bath. The solid line is the protocol leading to maximum power output $P^{*} = 0.043$ and corresponding efficiency $\eta^{*} = 0.29$. The dashed line corresponds to $P = 0.8 P^{*} = 0.034$ and $\tilde{\eta} = 0.39$. Remarkably, we have $\tilde{\eta}/\eta^{*} = 1.34$; the increase in efficiency is therefore larger than the reduction in power.  \label{fig:protocol}}
\end{figure}

 Figure \ref{fig:protocol} presents the optimal protocols \eqref{protocol-under-2} for fixed power $P_0$ and maximum power $P^*$. Their qualitative behavior may be easily understood.
During the coupling to the hot bath (red shaded area), the frequency $\omega$ of the harmonic potential is reduced  exponentially.
This step allows the particle to explore a larger space, similar to the expansion phase of a standard heat engine.
The continuous exponential expansion is followed by an instantaneous expansion step (jump)  immediately after the coupling to the cold bath. 
During the cold phase (blue shaded area), the  frequency increases exponentially, again followed by an instantaneous compression step.
This  behavior is the same irregardless of the power. However, the precise values $\omega_i$ of the trap frequency do depend on the power.
The discontinuities in the optimal protocols  also occur in the overdamped regime \cite{sch07}. They have their physical origin in the sudden  temperature change when switching  between the heat baths.

\begin{figure}[ht]
\includegraphics[width=0.48\textwidth]{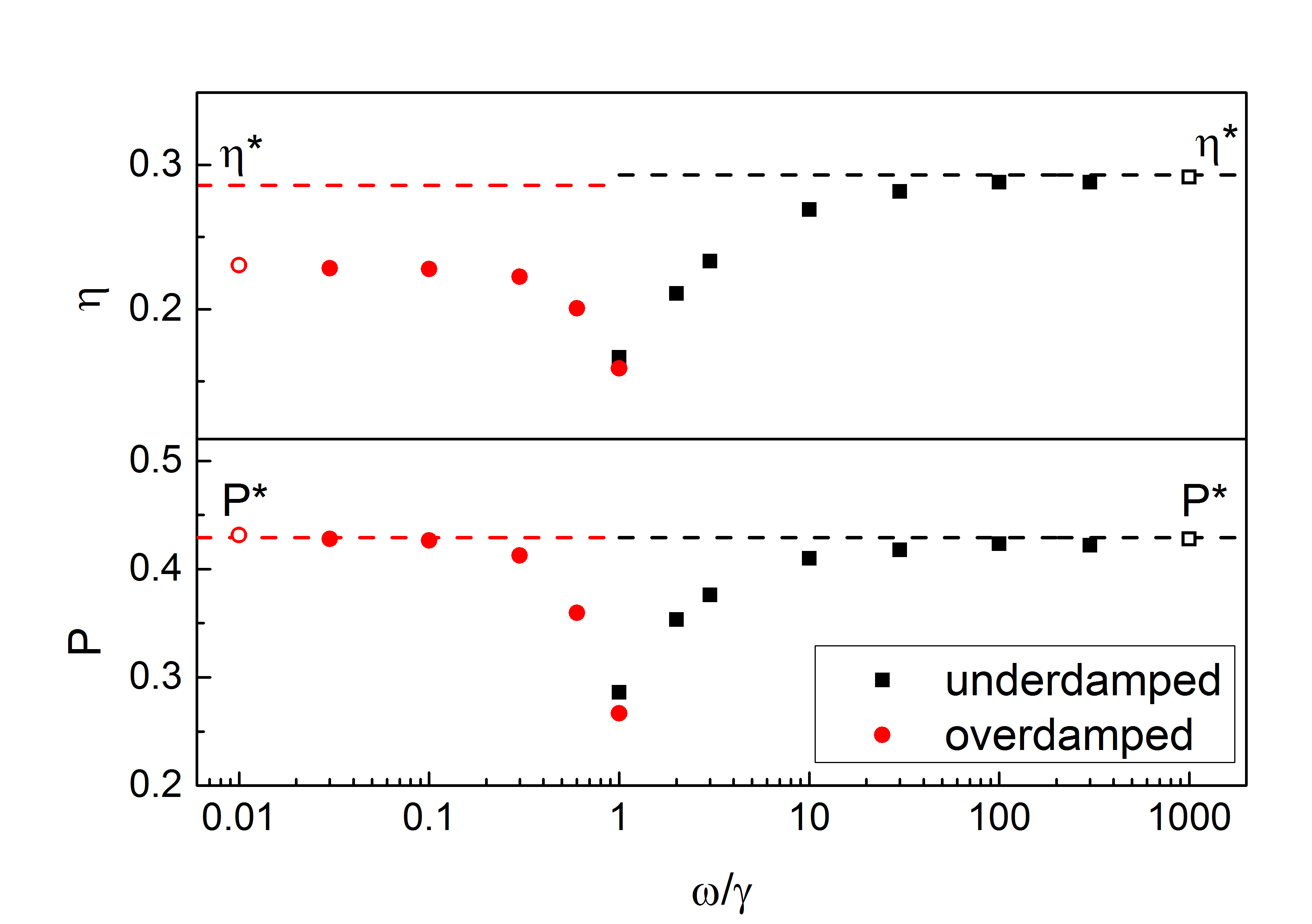}
\caption{(color online) Efficiency (top) and power (bottom) as a function of the ratio $\omega/\gamma$. The dashed lines indicate the theoretical results Eq.~\eqref{max-efficiency-under} for the efficiency at maximum power $\eta^{*}$ for $y=1$ and Eq.~\eqref{max-power-under} for the  maximum power $P^{*}$ (for the corresponding overdamped results, see Ref.~\cite{sch07} and Supplemental Material). The symbols are obtained by numerically solving Eq.~\eqref{variances} using the underdamped (black squares, Eq.~\eqref{underdamped-solution}) respectively overdamped (red circles, \cite{sch07}) optimal protocols. For the maximum power, the theoretical value is obtained both for the underdamped $\omega \gg \gamma$ and the overdamped $\omega \ll \gamma$ limit. While the efficiency is correctly predicted by Eq.~\eqref{max-efficiency-under} in the underdamped limit, the observed value in the overdamped case is lower than predicted in the usual overdamped limit  \cite{sch07}. \label{fig:maxpower}}
\end{figure}

\textit{Discussion.} Within the underdamped limit, the particle is able to adjust instantaneously to  changes in the trap frequency. This is of course an approximation for any real physical system.
In order to verify its validity, we numerically solve the full dynamics \eqref{variances}, for arbitrary $\gamma$, using the optimal protocols obtained in the underdamped limit, as shown in Fig.~\ref{fig:maxpower}.
For comparison, we also include the results from the overdamped case obtained by Schmiedl and Seifert \cite{sch07} (see also Supplemental Material).
In the underdamped regime, the optimal protocol \eqref{underdamped-solution} leads to  values of  efficiency and power that are close to Eqs.~\eqref{max-efficiency-under} and \eqref{max-power-under} for $\omega \gtrsim 10 \gamma$.
For any finite frequency $\omega$, the underdamped description \eqref{variance_v} requires that the particle is able to follow variations in the protocol. As a result, the rate of change of the frequency cannot be arbitrarily fast.
This leads to the condition $\dot{\omega}/\omega \ll \omega$, which we take into account by replacing the instantaneous jumps at $\tau_h$ and $\tau_h+\tau_c$ with a linear variation of length $\tau_\Delta$, such that $\dot{\omega}/\omega < \omega/10$. This provides results in good agreement with the analytical predictions.
As the ratio $\omega/\gamma$ decreases, both  power and efficiency start to drop as the underdamped approximation  is no longer valid.
For $\omega/\gamma < 1$, the overdamped optimal protocols yield higher power output which approaches the predicted maximum value for $\omega \lesssim 0.1 \gamma$.
The efficiency at maximum power for the full simulated dynamics, however, remains smaller than predicted in the overdamped limit, even at very small ratios of $\omega/\gamma$.
The reason for this deviation is the fact that the relaxation of the velocity, which is neglected in the overdamped definition of heat, leads to increased dissipation which does not vanish when taking the overdamped limit \cite{hon00,sch07}. In the intermediate regime, $\gamma \sim \omega$, where the coupling to the bath is neither weak nor strong, our numerical analysis shows that the optimal protocols derived in both underdamped and overdamped limit fail. The explicit optimal protocols are here not known due to the increased complexity of the optimization procedure \cite{gom08,mur14a,mur14}.

\textit{Experimental considerations.} 
Our theoretical findings might be verified using  a micromechanical harmonic oscillator with controllable spring constant and temperature. A possible candidate system in the underdamped regime is levitated cavity optomechanics \cite{rom10,cha10,bar10,kie13,mil15} where the parameters can essentially be tuned independently.The environement determines the inital temperature and Stokes friction for the particle. The effective temperature of the baths is then controlled via the intracavity light field and the spring constant via the power of the optical trap. The variation of the latter two parameters   is sufficiently fast to implement nearly optimal protocols \cite{dec15}. In the overdamped regime, however, only very small temperature differences are achievable. A possible trick to reach fast switching with a large temperature difference is to control the effective temperature via an artificially applied force noise \cite{mar15}. For the purpose of feedback cooling, external optical or electric drivings have been used in a variety of experiments with levitated nanoparticles \cite{ash77,li11,moo14,ran15,gra15}. An experimental demonstration may thus be readily implemented in these setups.

\textit{Conclusions.}
We have optimized an underdamped stochastic harmonic heat engine for maximum efficiency at fixed power. We have analytically determined the optimal driving protocols for all values of the power. They differ from those obtained in the overdamped limit,   but like the former exhibit jumps that mirror the sudden change in temperature. We have demonstrated that the Curzon-Ahlborn efficiency is the universal efficiency at maximum power due to the linear heat transport implicit in the weak coupling limit. We have further shown that 
reducing the power slightly below its maximum value leads to disproportionately large gains in efficiency.
Our findings confirm that maximizing the efficiency at maximum power is not always the best way to design machines with both high efficiency and high power. 

\begin{acknowledgments}
\textit{Acknowledgments.} This work was partially supported by the EU Collaborative Project TherMiQ (Grant Agreement 618074) and the COST Action MP1209. A.D. was partially employed as an International Research Fellow of the Japan Society for the Promotion of Science.
\end{acknowledgments}

\section*{Supplemental Material}

\subsection{Underdamped optimization}
The central insight for solving the Euler-Lagrange equations (8) in the main text comes from Eq.~(8b), which shows that $\alpha \sigma_v$ is constant.
Since neither $\alpha \equiv 0$ nor $\sigma_v \equiv 0$ are viable solutions, we may set $\alpha = -C/\sigma_v$ with some $C \neq 0$.
We then find from Eqs.~(8a) and (8c), $\sigma_v = \sqrt{C \gamma T/m}$, implying that $\sigma_v$ is constant.
The only way to have a constant $\sigma_v$ from Eq.~(8c) is with an exponential protocol $\omega(t)  = \omega_0 e^{\kappa t}$ with $\kappa = \gamma(1-T/(m \sigma_v))$.
Accordingly,
\begin{align}
\omega(t) = \left\lbrace \begin{array}{ll}
\omega_h e^{\gamma_h\big(1-\frac{T_h}{m \sigma_h}\big) t} &\text{for} \; 0 < t < \tau_h \\[2ex]
\omega_c e^{\gamma_c\big(1-\frac{T_c}{m \sigma_c}\big) (t-\tau_h)} &\text{for} \; 0 < t - \tau_h < \tau_c .
\end{array} \right. \label{protocol-under}
\end{align}
The exponential form of the protocols \eqref{protocol-under} has been  obtained in Ref.~\cite{aga13}.
However, the crucial point for the optimization of a periodically operating  engine, overlooked in Ref.~\cite{aga13}, is the implementation of  the proper boundary conditions.
The parameters $\omega_i$ and $\sigma_i$ are indeed not independent, but are fixed by the boundary conditions at $t = \tau_h$ and $t = \tau_h+\tau_c$, in contrast to the overdamped case (see below)
In principle, the protocol $\omega(t)$ can be continuous or discontinuous at these points.
Multiplying Eq.~(4) by $\omega$ and integrating the resulting equation from $\tau_h - \epsilon/2$ to $\tau_h + \epsilon/2$, with $\epsilon$  small, we find,
\begin{align}
\int_{\tau_h-\frac{\epsilon}{2}}^{\tau_h+\frac{\epsilon}{2}} \text{d}t \ \big[\dot{\sigma_v} \omega - \sigma_v \dot{\omega} \big] = \mathcal{O}(\delta) \label{boundary} .
\end{align}
If we allow  discontinuities, the frequency $\omega$ jumps from the value $\omega_{-} = \omega_h \exp(\gamma_h(1-T_h/(m \sigma_h))\tau_h)$ at $t = \tau_{h}^{-}$ to $\omega_{+} = \omega_c$ at $t = \tau_{h}^{+}$.
At the same time, $\sigma_v$ jumps from $\sigma_{-} = \sigma_h$ to $\sigma_{+} = \sigma_c$.
Close to $t = \tau_h$, we may thus write, 
\begin{align}
\omega(t) &= \omega_{-} + (\omega_{+}-\omega_{-}) \theta(t-\tau_h), 
 \\
 \Rightarrow \dot{\omega}(t) &= (\omega_{+}-\omega_{-}) \delta(t-\tau_h), \nonumber \end{align}
and similarly for $\sigma_v$.
Plugging the above expressions into Eq.~\eqref{boundary}, we obtain the condition $\sigma_{+} \omega_{-} = \sigma_{-} \omega_{+}$.
By contrast, if we demand that $\omega$ should be continuous, $\omega_{+}=\omega_{-}$,  the same has also to be true for $\sigma_v$.
However, if $\sigma_v$ stays constant during the entire cycle, both heat transfers are zero. 
This trivial solution is incompatible with the requirement of finite power. Hence maximization of the extractable
work requires a discontinuous protocol. From the condition $\sigma_{+} \omega_{-} = \sigma_{-} \omega_{+}$ and a similar one at $t = \tau_h + \tau_c$, we can eliminate $\omega_h$ from Eq.~\eqref{protocol-under}, find a relation between $\sigma_h$ and $\sigma_c$, and so obtain Eqs.~(9).

\subsection{Overdamped optimization}

In the overdamped limit, the maximum power and  the corresponding efficiency have been obtained in Refs.~\cite{sch07}.
The optimal protocols are, in terms of $\sigma_x$,
\begin{align}
\sigma_x(t) = \left\lbrace \begin{array}{ll}
\sigma_h \Big(1 + \frac{t}{\tau_h} \frac{\sigma_c - \sigma_h}{\sigma_h + \sqrt{\sigma_h \sigma_c}} \Big)^2 &\text{for} \; 0 < t <\tau_h \\[2ex]
\sigma_c \Big(1 + \frac{t-\tau_h}{\tau_c}\frac{\sigma_h - \sigma_c}{\sigma_c + \sqrt{\sigma_h \sigma_c}} \Big)^2  &\text{for} \; 0 < t-\tau_h <\tau_c . 
\end{array} \right. \label{protocol-overdamped}
\end{align}
Here $\sigma_x$ is a continuous function, whereas the resulting frequency protocol $\omega(t)$, determined from Eq.~(3), is generally discontinuous.
Maximizing the power with respect to $\tau_h$ and $\tau_c$ yields the equations \cite{sch07},
\begin{align}
P^{*} &= \frac{\ln^2 \big(\frac{\sigma_c}{\sigma_h}\big) (T_h - T_c)^2}{16 \big(\gamma_h + \gamma_c + 2 \sqrt{\gamma_h \gamma_c}\big)\big(\sigma_h + \sigma_c - 2 \sqrt{\sigma_h \sigma_c}\big)} \label{max-power-over} \\
\eta^{*} &= \frac{\Big(1+\sqrt{\frac{\gamma_c}{\gamma_h}}\Big) \Big(1-\frac{T_c}{T_h}\Big)}{2 \Big(1+\sqrt{\frac{\gamma_c}{\gamma_h}}\Big)-\Big(1-\frac{T_c}{T_h}\Big)}. \label{max-power-efficiency-over}
\end{align}
Contrary to the underdamped limit, there exist no optimal values for $\sigma_h$ and $\sigma_c$.
This can be seen from Eq.~\eqref{max-power-over}; the maximum power diverges if one of the two parameters tends to zero while the other one stays constant.
Physically, this limit corresponds to having a very large frequency, at which point the overdamped limit is no longer valid.
Thus within the overdamped approximation, $\sigma_h$ and $\sigma_c$ are free parameters that have to be chosen such that the condition $\omega \ll \gamma$ remains valid.

Comparing the maximal values for the power Eqs.~(12) and \eqref{max-power-over}, we see that in the underdamped limit, the power is proportional to the overall damping strength, whereas it is inversely proportional in the overdamped regime.
Since the damping is assumed to be small in the underdamped limit and large in the overdamped one, the power itself is small in both limits, owing to the asymptotic nature of the treatment.
Physically, the small resultant power has different reasons: 
In the underdamped limit, the particle exchanges energy with the bath at a small rate, limiting the total transferred energy per cycle.
In the overdamped limit, the energy exchange with the bath is fast, however the reaction to changes in the external potential is slow.


\begin{thebibliography}{99}
\bibitem{com} In the overdamped limit, Schmiedl and Seifert have shown that the efficiency at maximum power  exceeds $\eta_{CA}$ when the mobility (i.e. the inverse damping coefficient) decreases faster than $1/T$ \cite{sch07}.
\end{thebibliography}
\end{document}